\documentclass[a4paper,11pt]{article}
\pdfoutput=1 

\usepackage{jcappub} 

\usepackage[T1]{fontenc} 
\usepackage[version=4]{mhchem}
\usepackage{upgreek}
\usepackage{footnote}
\makesavenoteenv{tabular}
\usepackage{tablefootnote}

\title{\boldmath Echo-free quality factor of a multilayer axion haloscope}

\author[a, b]{Juan F. Hernández-Cabrera,}
\author[c, a, b]{Javier De Miguel,}
\author[a]{E. Hernández-Suárez,} 
\author[a]{Enrique Joven,} 
\author[a]{H. Lorenzo-Hernández,} 
\author[c]{Chiko Otani,}
\author[a, b]{J. Alberto Rubiño-Martín} 
\author[d]{and Konstantin Zioutas} 
\author{on behalf of the DALI Collaboration}

\affiliation[a]{Instituto de Astrofísica de Canarias, \\E-38200 La Laguna, Tenerife, Spain}
\affiliation[b]{Departamento de Astrofísica, Universidad de La Laguna,\\ E-38206 La Laguna, Tenerife, Spain}
\affiliation[c]{The Institute of Physical and Chemical Research (RIKEN), Center for Advanced Photonics,\\ 519-1399 Aramaki-Aoba, Aoba-ku, Sendai, Miyagi 980-0845, Japan}
\affiliation[d]{Physics Department, University of Patras,\\ GR 26504, Patras-Rio, Greece}

\emailAdd{javier.miguelhernandez@riken.jp}

\abstract{We report a methodology to determine the quality factor ($Q$) in implementations of the so-called dielectric haloscope, a new concept of wavy dark matter detector equipped with a multilayered resonator. An anechoic chamber enables the observation of the resonance frequency and its amplitude for an  unlimited series of layers for the first time, which is conveniently filtered. The frequency-normalized power enhancement measured in a Dark-photons \& Axion-Like particles Interferometer (DALI) prototype is a few hundred per layer over a sweep bandwidth of half a hundred MHz. In light of this result, this scaled-down prototype is sensitive to axions saturating the local dark matter density with a coupling to photons between $g_{a\gamma\gamma}\gtrsim10^{-12}$ GeV$^{-1}$ and $g_{a\gamma\gamma}\gtrsim \mathrm{few} \times 10^{-14}$ GeV$^{-1}$ at frequencies of several dozens of GHz once cooled down to the different working temperatures of the experiment and immersed in magnetic fields ranging from 1 T to 10 T; while the sensitivity of the full-scale DALI is projected at $g_{a\gamma\gamma}\gtrsim\mathrm{few}\times10^{-15}$ GeV$^{-1}$ over the entire 25--250 $\upmu$eV range since $Q\gtrsim10^4$ is expected.}

\begin{document}
\maketitle
\flushbottom

\section{Introduction}
\label{sec:intro}

\begin{figure}[tbp]
    \centering    \includegraphics[width=0.65\textwidth]{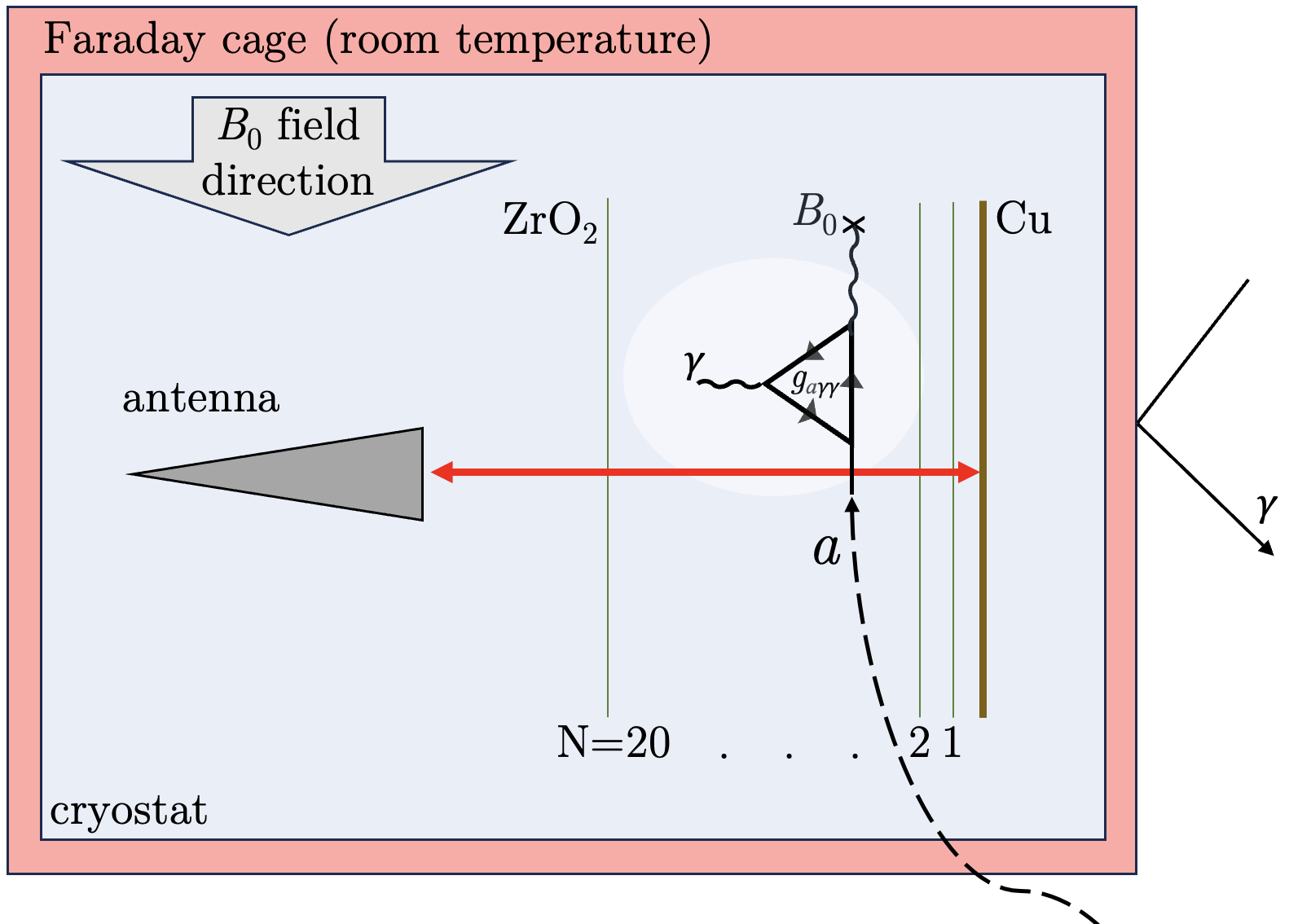}
    \caption{Working principle of the 20-layer, 1-pixel, scaled-down DALI prototype. The experiment is housed in a Faraday cage to attenuate microwave signals from the outside to the inside. Therefore, spurious photons ($\gamma$) are prevented to access the vessel while the dark matter axions ($a$) pass through it. The cryostat is magnetized. A part of the ambient axions are converted into microwaves that are received by an antenna and sent to the acquisition system (not shown). The Fabry-Pérot resonator, composed of 20 layers of zirconium oxide and wrapped by a copper mirror, renders the weak signal induced by the ALPs in our galactic halo observable in a reasonable integration time.  }
    \label{fig_0}
\end{figure}

The axion is a long-postulated pseudoscalar boson that emerges from the quantum chromodynamics (QCD) solution to the intriguing conservation of charge and parity (\textit{CP}) in the strong nuclear force \cite{PhysRevLett.38.1440, PhysRevLett.40.223, PhysRevLett.40.279}. The axion-based solution to the \textit{CP} problem is compatible with the non-observation of the neutron electric dipole moment in experiments \cite{PhysRevLett.124.081803, PhysRevLett.97.131801}. Moreover, due to its lightness and weak interaction with baryonic matter, the axion is a promising candidate to make up, in part or in whole, the so-called dark matter (DM), an unknown substance  widely assumed to be non-luminous \cite{1933AcHPh...6..110Z, ABBOTT1983133, DINE1983137, PRESKILL1983127}. Although there is indirect evidence of DM, it has never been directly detected. The axion has been searched for with various experimental-theoretical approaches, including Refs. \cite{PhysRevD.105.035022, PhysRevD.98.103015, Marsh_2017, PhysRevLett.118.011103, 2014PhRvL.113s1302A, Straniero:2015nvc, 2022JCAP...10..096D, 2022JCAP...02..035D, REGIS2021136075, HAYSTAC:2020kwv, EHRET2010149, PhysRevD.88.075014, EJLLI20201,  PhysRevD.92.092002,PhysRevLett.59.839, PhysRevD.42.1297, PhysRevLett.104.041301,PhysRevD.97.092001,PhysRevLett.128.241805,doi:10.1063/5.0098783, CAST:2017uph, CAST:2020rlf, Foster:2020pgt, Darling:2020uyo, Darling:2020plz, PhysRevLett.121.261302, PhysRevD.99.101101, PhysRevD.103.102004, MCALLISTER201767, doi:10.1126/sciadv.abq3765, AxionLimits, Quiskamp:2023ehr}. In this paper we focus on an experimental direct search for Galactic DM proposed by Sikivie in the 1980s, the axion haloscope~\cite{PhysRevLett.51.1415}. Encouraged by the high density of DM haloes in common spiral galaxies like the Milky Way \cite{1970ApJ...159..379R}, the haloscope was envisioned to probe DM at the position of a laboratory on Earth. In the most widely used setup to date, ambient axions are converted into microwaves in a magnetized resonant cavity via the inverse Primakoff effect \cite{Primakoff:1951iae}. However, the cavity haloscope presents some technical hurdles for masses of several dozen $\upmu$eV and beyond. In particular, the miniaturization that accompanies an increase in frequency progressively reduces the magnetized volume, which in turn reduces the power of the output signal, drastically lowering sensitivity. This is particularly disadvantageous because the exploration of the sector corresponding to heavier axions is well motivated by inflationary cosmology in the band, say, from 10 to 50 GHz  \cite{2022NatCo..13.1049B}. This technical challenge has led to the search for alternative haloscope concepts that will allow us to probe DM axions at high frequency, such as the so-called dielectric haloscope \cite{MADMAXinterestGroup:2017koy}. Evolved from the dish-antenna setup \cite{Horns:2012jf}, this experimental approach employs a Fabry-Pérot interferometer to enhance the weak axion-induced signal rather than a resonant cavity \cite{1899ApJ.....9...87P}. Crucially, the scanning frequency is uncoupled from the area of the dielectric plates, provided it is larger than the signal wavelength to avoid diffraction, allowing for a high sensitivity at higher frequencies. Different interpretations of this design are under research and development in diverse locations and mass ranges, including ADMX-Orpheus \cite{PhysRevD.106.102002, PhysRevLett.129.201301}, DALI \cite{De_Miguel_2021, DeMiguel:2023nmz}, DBAS  \cite{PhysRevApplied.9.014028, PhysRevApplied.14.044051}, LAMPOST \cite{PhysRevD.98.035006, Chiles:2021gxk}, MADMAX \cite{MADMAXinterestGroup:2017koy, MADMAX:2019pub}, and MuDHI \cite{ PhysRevD.105.052010}. We focus on DALI throughout this work, which introduces new features that no other haloscope has ever addressed before—c.f. \cite{ DeMiguel:2023nmz, Cabrera:2023qkt}. First, DALI pioneers a steerable mount to boost the directional search for DM substructures. Second, DALI incorporates a multipixel focal plane which, transferred from our expertise in radio astronomy \cite{1994Natur.367..333H, 10.1046/j.1365-8711.2003.06338.x, 2011A&A...536A...1P, 10.1093/mnras/stac3439, 2020SPIE11453E..0TR}, allows DALI to simultaneously scan multiple resonant frequencies that are periodic every four wavelengths—e.g. $\sim$$\lambda/8$ for a one-eighth wavelength stack and $\sim$$\lambda/2$ for a half-wavelength stack—thus saving room for plate spacing while doubling the scanning speed. Another key factor of the DALI experiment is that it employs only available equipment. As a sample, the use of a solenoid-type superconducting magnet such as those used in medical resonance imaging makes the experiment feasible while reducing its cost compared to other proposals to probe axions. Like some other axion haloscopes, DALI is also responsive to other wavy dark matters such as the dark photon \cite{Okun:1982xi} and the so-called axion-like particles (ALPs), in which mass and coupling constant are decoupled \cite{doi:10.1146/annurev-nucl-120720-031147}; and has the capacity to transform gravitons into photons within its magnetized vessel allowing to set new constraints at high frequency to a gravitational wave background \cite{DeMiguel:2023nmz}. The~experiment, currently in a prototyping phase, is to~be installed at the Teide Observatory, in~the Canary Islands, in an environment protected from terrestrial microwave sources. The DALI program comprises different projects. As part of its scientific plan, in this work we forecast the sensitivity of a proof-of-principle prototype of DALI, at a lower scale, designed to retain a part of its physics potential but at the same time relieving some hardware heaviness by halving the number of layers of the Fabry-Pérot tuner, from 40--50 to 20; and scaling the plate size to $\sim$1:10, which diminishes the detection cross-sectional area accordingly, while the data stream is smoothed by adjusting the number of channels to a single pixel. All of the above attenuates the sensitivity of the prototype respect to the full-scale DALI but, interestingly, the scaling still preserves a remarkable discovery capability, as we will show in this manuscript. The tuner concept was presented in \cite{2024JInst..19P1022H}. A scissor mechanism, similar to an accordion, forces consecutive layers of the Fabry--Pérot resonator to move while maintaining the same plate spacing, allowing for frequency tuning and simultaneously restricting unwanted degrees of freedom. The operational ranges of the prototype are centered at $\sim$7 GHz with a few GHz bandwidth and, in a second stage, centered at $\sim$33 GHz with several GHz bandwidth in search for axion-like and dark photon DM. A~scheme is shown in Fig. \ref{fig_0}. The axion-induced signal power from the interferometer is

\begin{equation}
P \sim 10^{-25}\,\mathrm{W} \times\frac{A}{\mathrm{0.01\,m^2}}\times\left(\frac{B_0}{1\,\mathrm{T}}\right)^2\times Q \times \left (\frac{g_{a\gamma\gamma}}{10^{-12} \,\mathrm{GeV}^{-1}} \right )^2 \times\left(\frac{10\,\upmu \mathrm{eV}}{m_a}\right)^2\frac{\rho_a}{0.4\,\mathrm{GeVcm^{-3}}}\;,
\label{Eq.1}
\end{equation}

 where $A$ is the cross section area, $B_0$ is the external magnetic field, $g_{a\gamma\gamma}$ is the axion--photon coupling, $m_a$ is the axion mass, and $\rho_{a}$ is the local density of axion DM—e.g., see \cite{Staudt:2024tdq}. In particular, the estimation of the quality factor, $Q$, a figure of power enhancement around the resonant frequency, presents some challenges associated with reflections, and had never been achieved for haloscopes longer than a few layers before this work \cite{Egge:2020hyo, 2024JInst..19P1022H}. In \cite{Egge:2022gfp}, a promising method based on the reciprocity of reflections that may be used for $Q$ determination is discussed. In this article, we report the first direct observation of $Q$ for a multilayer haloscope, a dummy of DALI, in an anechoic chamber that cancels out undesired reflections. The rest of the article is structured as follows. In Sec. \ref{Sec2} we describe the methodology for the measurement of $Q$. In Sec. \ref{Sec3} we discuss the results, while some relevant conclusions are drawn in Sec. \ref{Sec4}.

\section{Anechoic measurement of the quality factor}
\label{Sec2}
To conduct a measurement of the $Q$ factor, two fixed-plate Fabry-Pérot resonators holding $\mathrm{N} = 4, \ldots, 20$ yttria-stabilized zirconia (YSZ) layers \cite{physrevb.65.075105, physrevb.64.134301} with dimensions $(100 \times 100) \pm 0.5$ mm and a thickness of $1 \pm 0.03$ mm have been manufactured with a measured error of the order of a few dozen microns or less in the spacing of the layers—6.04 mm and 6.21 mm—, whose surface roughness is lower than 0.1 $\upmu$m. At $\mathcal{O} (10)$ GHz frequencies the relative permittivity, $\varepsilon_{\mathrm{r}}$, of the tetragonal crystal phase structure of zirconia predominant in the layers given the yttria concentration of 3 mol\% \cite{Götsch_2016, Cho_2023} is in the range of 30 to 46 \cite{10.1063/1.353963, Lanagan_1989, Miranda1990, Tonkin_1997}, and loss tangents of the order of $\tan \delta \sim 10^{-4}$ are reported \cite{Centeno_1995, Iijima_1992, Smith_1992}. 

\begin{figure}[tbp]
    \centering 
    \includegraphics[width=0.63\textwidth]{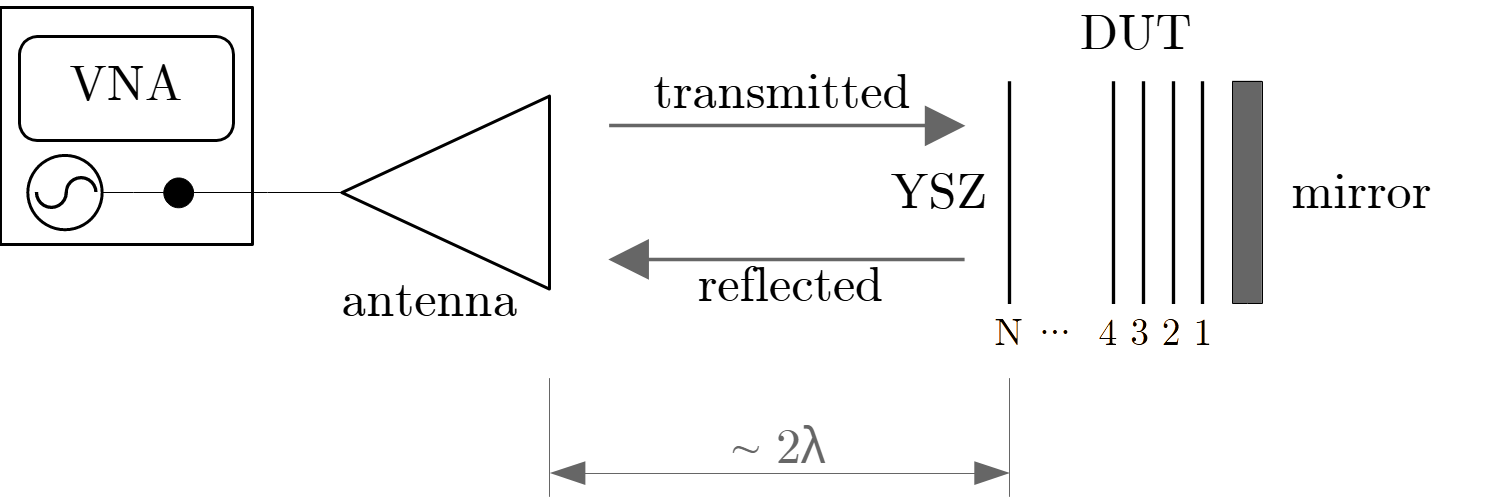}
    \caption{Schematic representation of the experimental setup for the measurement of the quality factor of the Fabry-Pérot resonators. A horn antenna placed at a distance of about $2 \lambda$---beyond the lower limit of the radiative near field---excites the resonator with a signal generated by a VNA, which traverses the YSZ plates and bounces off the bottom mirror; a standing wave is created and picked up by the antenna, which is registered by the VNA as a reflectivity scattering parameter.}
    \label{fig_1}
\end{figure}

A room-temperature experimental setup in a shielded microwave anechoic chamber\footnote{EMC SHC-BOX-L anechoic chamber by EMCTEST.} has been developed to reliably infer the quality factor of the two resonators as devices under test (DUTs) from measurements of scattering parameters in the microwave C band, which contains the resonant frequency $\nu_0$ corresponding to the one-eighth wavelength stacking of the resonators \cite{DeMiguel:2023nmz}. A vector network analyzer (VNA) was used to measure the reflectivity ($\Gamma$) of the DUT—$S_{11}$ scattering parameter—with a resolution of $\sim$ 300 kHz in an optical assembly with a pointing error of the order of a few degrees. The group delay, $\tau_{\mathrm{g}}$, is numerically obtained from the reflectivity as $\tau_{\mathrm{g}} = - \mathrm{d} / \mathrm{d} \omega \arg (\Gamma)$, with $\omega$ as the angular frequency, which is used to calculate the quality factor as $Q = \tau_{\mathrm{g}} \omega_0$ \cite{renk2012basics}. Spurious reflections off objects other than the resonator and the antenna in the vicinity of the setup are significantly attenuated in the anechoic chamber, which are further reduced by covering other reflective elements of the setup with additional pyramidal microwave absorbers. As a result, the response from the resonator and the cutoff of the antenna become the only significant reflections in the same time-domain range, which renders isolating the response of the DUT through time-domain windowing feasible. This eliminates the restriction on the measurable number of layers reported in previous similar experiments performed in Faraday cages, where strong reflections were present in the same time-domain range as the DUT response \cite{2024JInst..19P1022H, Hernandez-Cabrera:2024SPIE}. The setup is schematically represented in Fig. \ref{fig_1}. A flat copper mirror is added behind the last dielectric layer with the same spacing, ensuring phase coherence; frequency-swept reflectivity data is recorded for different $\mathrm{N}$. A horn antenna pointed toward the resonator transmits the excitation signal generated in the VNA and picks up the response of the DUTs, which are placed further than two wavelengths away from the outer rim of the antenna. The ripple originated by the positioning in the near field obscures small $\tau_{\mathrm{g}}$ peaks---i.e. originated by $\mathrm{N} \lesssim 4$---but does not impede the identification of larger peaks beyond an uncertainty in their heights of the order of a few nanoseconds. Attenuation of the response signal bounced off the DUTs---and therefore of its measured reflectivity---is irrelevant to the calculation of the quality factors. These graphs contain as many peaks as there are layers in the frequency band corresponding to a given configuration; the last one, which is the highest, determines the quality factor and the resonant frequency \cite{De_Miguel_2021}.

\begin{figure}[tbp]
    \centering 
    \includegraphics[width=0.83\textwidth]{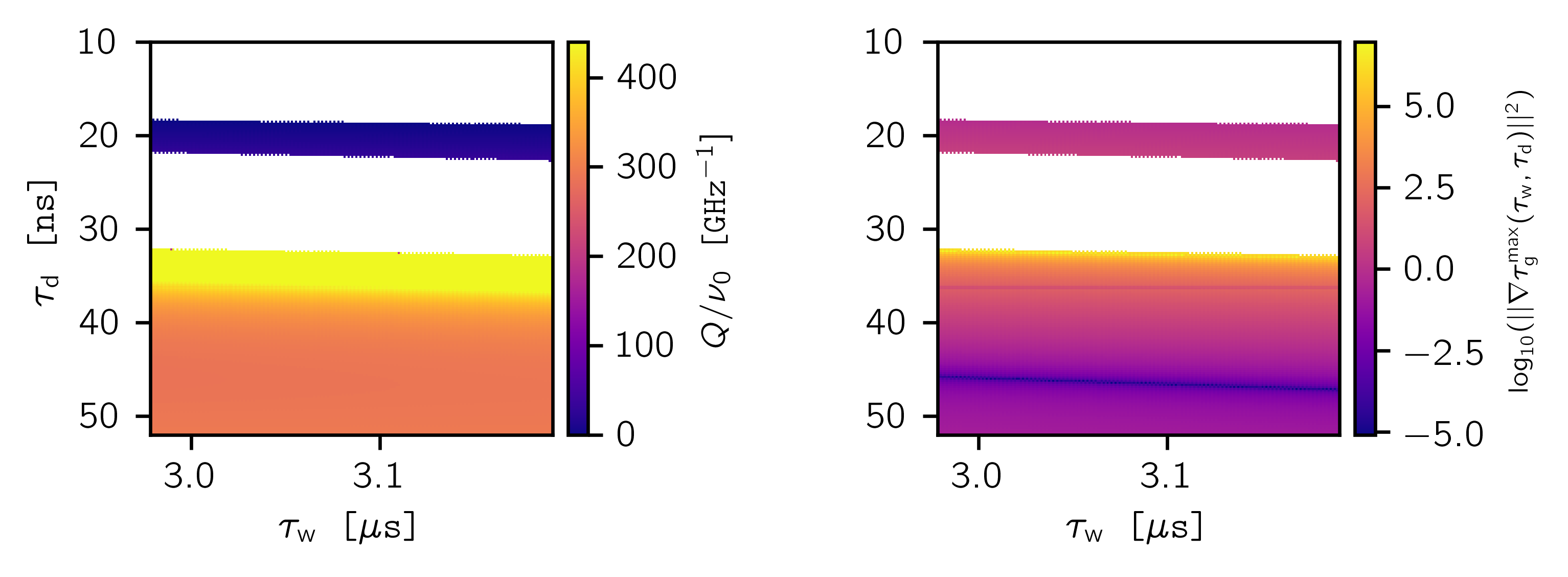}
    \caption{Normalized quality factor (left) and group delay gradient values of the DUT1 at $\mathrm{N}=20$ as functions of the $\tau_{\mathrm{w}}$---width---and  $\tau_{\mathrm{d}}$---location---parameters of a discrete prolate spheroidal sequence (DPSS) window, which is used to isolate the DUT response by time-domain gating. The lowest group delay gradient---$||\nabla \tau^\mathrm{max}_{\mathrm{g}} (\tau_{\mathrm{w}}, \tau_{\mathrm{d}})||^2$---is used to select the actual $Q$ value. White regions represent discarded negative and off-center $Q$ values.}
    \label{fig_2}
\end{figure}

Spurious reflections are reduced by using a discrete prolate spheroidal sequence (DPSS) window \cite{Slepian_1978} to apply time-domain gating, a technique which has been extensively used to isolate the response of a DUT in free-space radio frequency measurements \cite{Gonçalves_2018, Tang_2015, Baena_2015}. Reflectivity data is divided by the measured reflectivity of an empty resonator---mirror only---in order to eliminate systematic reflections such as the antenna cutoff, which is present in the same time scale as the DUT reflection, and to remove the group delay offset originated by the distance from the antenna radiating element to the DUT. The parameters of the DPSS window---width $\tau_{\mathrm{w}}$ and time-domain delay $\tau_{\mathrm{d}}$---for which $Q$ is recorded given a layer count $\mathrm{N}$ are selected such that the gradient $||\nabla \tau^\mathrm{max}_{\mathrm{g}} (\tau_{\mathrm{w}}, \tau_{\mathrm{d}})||^2$ of the maximum value of $\tau_{\mathrm{g}} (\omega)$ in a 20 kHz scan band centered in the group delay peak---represented as $\tau^\mathrm{max}_{\mathrm{g}}$---is minimized, with the goal of identifying a stable value of $Q$—see Fig. \ref{fig_2}. The DPSS half-bandwidth parameter is fixed at the same value as the window width. Negative values of $Q$ are eliminated from the process, as are peak values found at the edges of the scan band. The limits of the $(\tau_{\mathrm{w}}, \tau_{\mathrm{d}})$ parameter space were chosen so that $Q$ values with no physical meaning were discarded. We have normalized the quality factor to the resonant frequency in Table \ref{table_1} in order to show a frequency-decoupled figure of merit.

\section{Results and discussion}
\label{Sec3}
\begin{figure}[tbp]
    \centering 
    \includegraphics[width=0.61\textwidth]{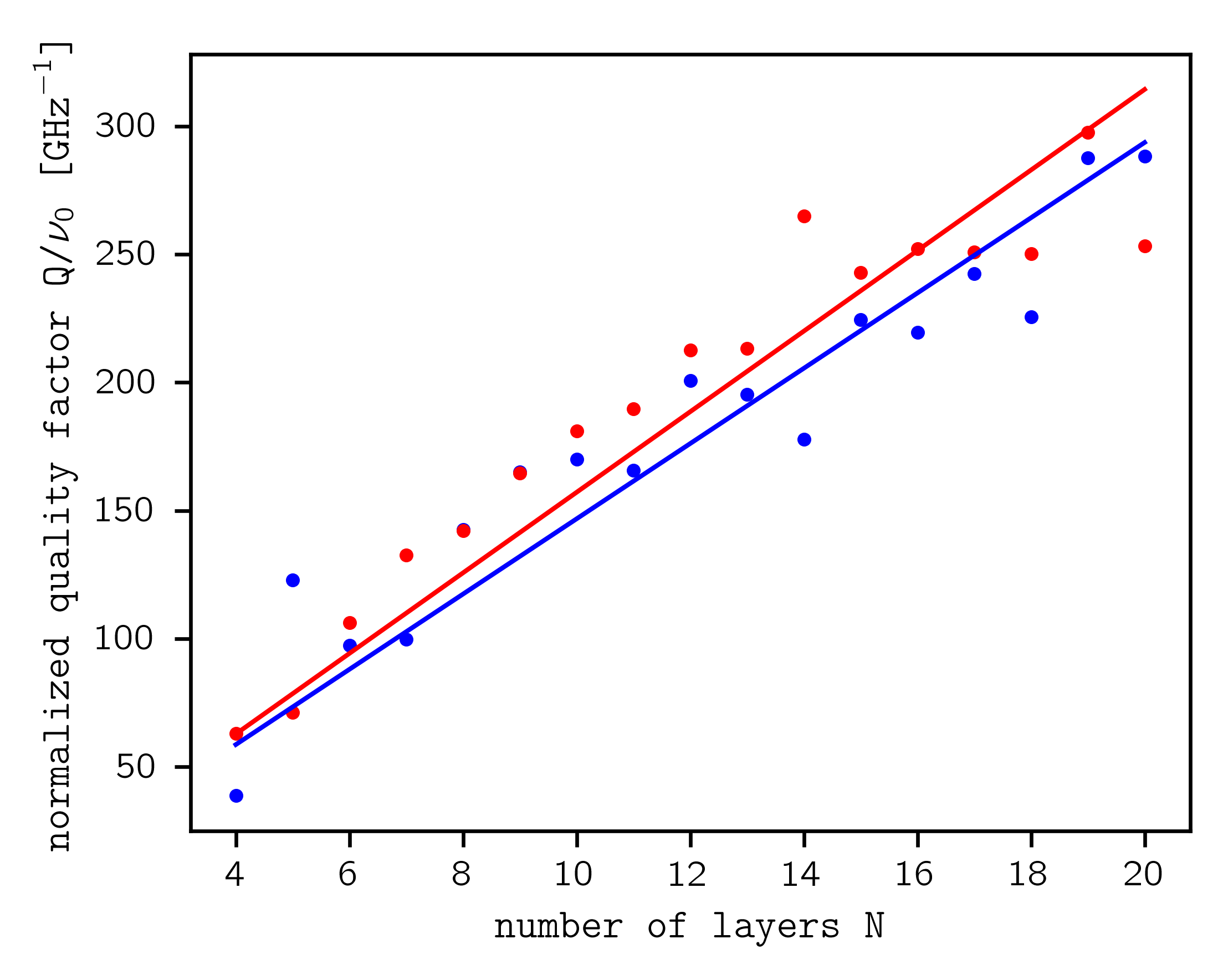}
    \caption{Measured quality factors of two fixed-plate dielectric haloscopes holding $\mathrm{N}$ layers with dimensions $100 \times 100 \times 1$ mm constructed with 3 mol\% yttria-stabilized zirconia with two different layer spacings, normalized to the resonant frequency, $\nu_0$. The measurements of DUT1 are shown in blue; those of DUT2 in red. The continuous lines represent the proportional fit reported in Table \ref{table_1}.}
    \label{fig_4}
\end{figure}

The measured quality factors of DUT1—6.04 mm spacing—and DUT2—6.21 mm spacing—are plotted against layer count $\mathrm{N}$ in Fig. \ref{fig_4}. At $\mathrm{N}=20$, resonance with a quality factor $Q \sim 2 \times 10^3$ was found at 7.01 GHz (DUT1) and 6.90 GHz (DUT2), a frequency shift expected from their spacings if also the effect of the layer thickness being different from the ideal $\sim \lambda / (8 \sqrt{\varepsilon_{\mathrm{r}}})$ is taken into account. A strong statistical link—coefficient of determination $r^2 \approx$ 0.9—between the layer count and the quality factor is found through a zero-intercept linear regression fit, which we report in Table \ref{table_1}. This is consistent with the behavior reported in existing literature \cite{Millar:2016cjp, millar2018theoretical}. 

\begin{table}[tbp]
\centering
\begin{tabular}{|c|c|c|c|}
\hline
DUT&$Q / (\nu_0 \mathrm{N}) \, [\mathrm{GHz}^{-1}]$&$r^2$&$\mathrm{RMSE}$\\
\hline
1 & 14.68 & 0.882 & 22.38\\
2 & 15.71 & 0.872 & 24.44\\
\hline
\end{tabular}
\caption{\label{table_1} Results of the zero-intercept linear regression fit between the quality factors and the layer count $\mathrm{N}$. The goodness of fit is reported as a coefficient of determination $r^2$ and $\mathrm{RMSE}$, which stands for the root mean square (RMS) error between the measured $Q / \nu_0$ and the predicted $\hat{Q} / \nu_0$. A RMS error around 8\% is found at a layer count of $\mathrm{N} = 20$.}
\end{table}

\begin{figure}[tbp]
    \centering 
    \includegraphics[width=0.58\textwidth]{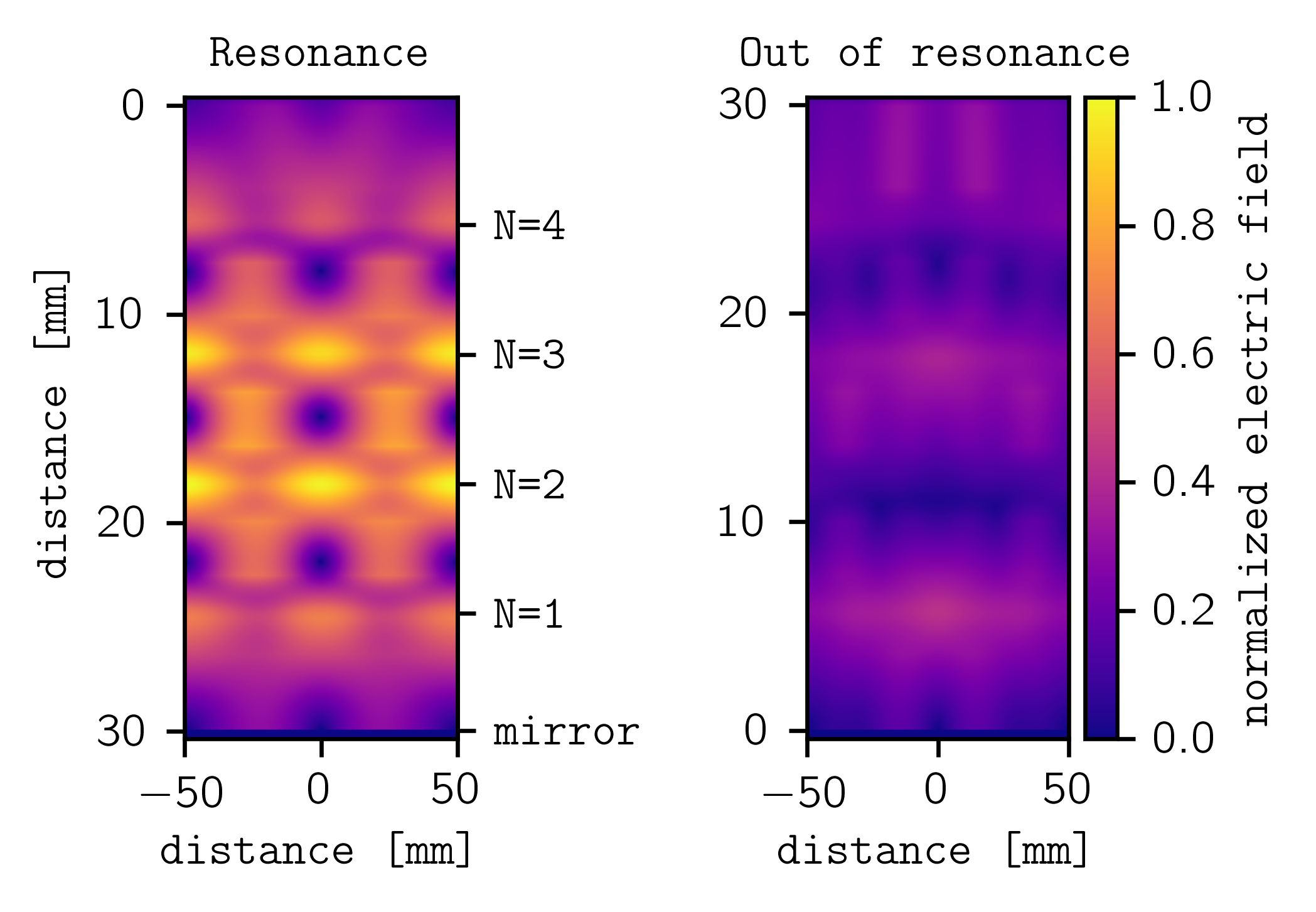}
    \caption{A finite element method simulation of the normalized electric field plotted on the transversal plane of a four-layer Fabry-Pérot interferometer at the $\lambda/8$ resonant frequency (left) and out of resonance (right). The wavelength is $\lambda\approx4$ cm. The locations of the layers ($\mathrm{N} = 1, \ldots, 4$) and the mirror are shown.}
    \label{fig_3}
\end{figure}
We have computed the electric field distribution with  electromagnetic finite element method simulations\footnote{This simulation was performed with CST Studio Suite\textsuperscript{\textregistered} v. 2021.00 licensed by IAC.} at the $\lambda/8$ resonant frequency and out of resonance, as shown in Fig. \ref{fig_3}. 

Resonance is expected to slightly differ in location and amplitude in a cryogenic environment, as the permittivity of 3 mol\% YSZ decreases with temperature according to \cite{Tonkin_1997}

\begin{equation}
\varepsilon_{\mathrm{r}} = 33.9 \left ( -2.01 \times 10^{-9} T^3 +  1.08 \times 10^{-6} T^2 +  8.93 \times 10^{-6} T + 0.96\right ) \,,
\label{Eq3.1}
\end{equation}

which is expected to bring about a small decrease in the quality factor per layer of a few percent given that $Q / (\nu_0 \mathrm{N})$ scales with $\varepsilon_{\mathrm{r}}$ \cite{Millar:2016cjp, millar2018theoretical}. However, the effect of dielectric loss tangent on the linearity of the dependence between $Q$ and $\mathrm{N}$ has been simulated in Refs. \cite{Millar:2016cjp, millar2018theoretical}, appearing to be insignificant at $\mathrm{N} = 80$ even for far lossier dielectrics with $\tan \delta \gg 10^{-4}$.  Moreover, it has been observed that the YSZ loss tangent decreases at liquid nitrogen temperatures as $\tan \delta = 10^{-5} (9.3 + 0.27 T)$ \cite{Smith_1992}, which would roughly compensate for the aforementioned loss of quality factor after cooling to the operating temperature of the experiment. Lastly, $\varepsilon_{\mathrm{r}}$ has been found to remain roughly constant from cm to mm wavelengths in earlier works \cite{10.1063/1.353963, Miranda1990, Hilario2019, Shimizu2022}. From all of the above, it follows that $Q \sim 10^4$ is tenable for $\mathrm{N} \approx 50$ at a few dozen GHz. Some benchmark values are shown in Table \ref{table_2}.

\begin{table}[h]
\centering
\begin{tabular}{|c|c|c|c|c|}
\hline
$\nu_0$ [GHz]& N & $T$ [K] & $Q_\mathrm{DUT 1}$ & $Q_\mathrm{DUT 2}$ \\
\hline
 & 20 & 290 & 2054\textsuperscript{*} & 2198\textsuperscript{*} \\
7 & 20 & 25 & 1965 & 2103\\
 & 50 & 0.2 & 4908 & 5253\\
\hline
 & 20 & 290 & 9685 & 10364\\
33 & 20 & 25 & 9264 & 9914\\
 & 50 & 0.2 & 23139 & 24763\\
\hline
 & 20 & 290 & 17608 & 18844\\
60 & 20 & 25 & 16844 & 18026\\
 & 50 & 0.2 & 42071 & 45023\\
\hline
\end{tabular}
\caption{\label{table_2} Reported and expected values of the quality factor. The three resonance frequencies, at 7 GHz, 33 GHz and 60 GHz, correspond to the prototype and full-scale DALI. The dummy tuner has $\mathrm{N}=20$ ZrO$_2$ layers, while the full-scale interferometer will have at least 50. Values for room-temperature measurements, a dielectric plate temperature of some 25 K achieved by means of $^4$He cryogenics, and about 0.2 K with $^3$He dilutors, are considered. Both DUTs, manufactured to resonate some 100 MHz apart in the lower frequency range, are shown separately. The superscript * denotes values as obtained from the observations—see linear fits in Table \ref{table_1}—, while the rest are estimated from the dielectric properties modified with temperature according to Eq. \ref{Eq3.1}. Maximal values of $Q\sim 5\times10^4$ are tenable in the experiments.}
\end{table}

\begin{figure}[h]
    \centering    \includegraphics[width=0.64\textwidth]{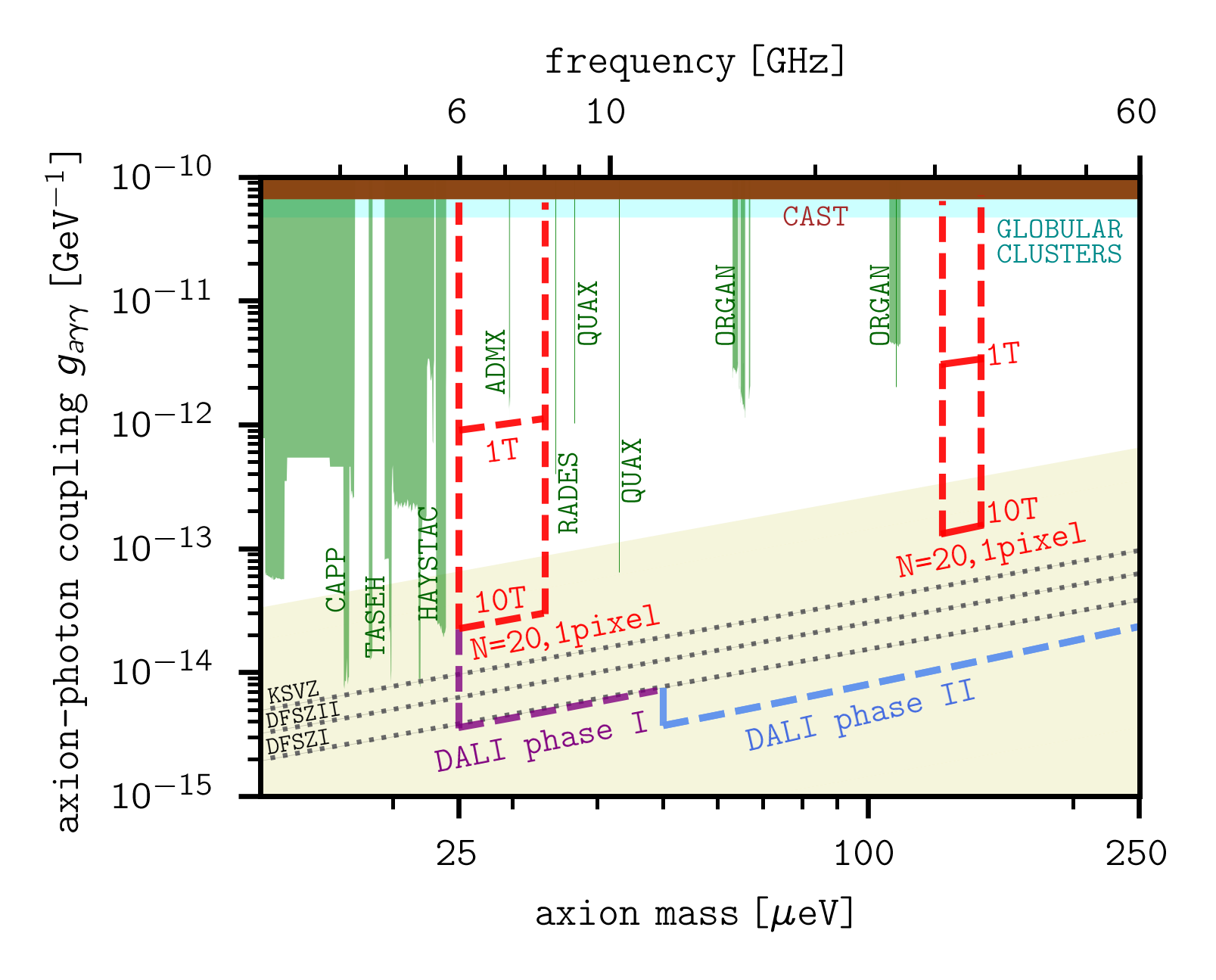}
    \caption{The sensitivity projections of the DALI experiments from Eq. \ref{Eq.2}. The parameters of the DALI prototype scaled to a single pixel, in red, are: $\mathrm{N}=20$, $\mathrm{SNR}=3$, $Q$ as reported in Sec. \ref{Sec3}, $A=0.01$ m$^2$, $t=12$ h, $T_{\mathrm{sys}}\approx30$ K with a physical temperature powered by a closed-cycle $^4$He system, $B_0=1$ T and $B_0=10$ T with a system temperature of $T_{\mathrm{sys}}\approx 1 + 3\times h\nu_0/(k_{\mathrm{B}}T)$ K reached by means of $^3$He cryogenics, $k_{\mathrm{B}}$ being the Boltzmann constant. The projections for both phases of the full-scale DALI are adapted from Ref. \cite{DeMiguel:2023nmz}—$A$ is 1/2 m$^2$ or 3/2 m$^2$, $B_0$ is 9.4 T or 11.7 T, $\mathrm{N}=50$, etc. The axion density at the Solar System position is fixed at $\rho_a=0.45$ GeV cm$^{-3}$. Current exclusion limits in the mass-coupling plane and reference axion models are shown. The \textit{phenomenologically preferred} axion window is shaded in beige \cite{DiLuzio:2016sbl}. Some reference axion models are included \cite{PhysRevLett.43.103, Shifman1980CanCE, DINE1981199, osti_7063072}.}
    \label{fig_5}
\end{figure}

It may be useful to round out the discussion with a simulation in light of these results. The sensitivity of the DALI instruments to an isotropic distribution of virialized DM axions reads \cite{DeMiguel:2023nmz, Cabrera:2023qkt} 
\begin{equation}
\begin{aligned}
g_{a \gamma\gamma} \gtrsim 5.4\times10^{-14} \, \mathrm{GeV^{-1}}  \times \left(\frac{\mathrm{SNR}}{5}\right)^{1/2}  \!\! 
\times \left(\frac{\mathrm{10^4}}{Q}\right)^{1/2}  \!\! \times 
\left(\frac{\mathrm{1\,m}^2}{A}\right)^{1/2}\!\! \times  \left(\frac{m_{a}}{\mathrm{100\,\upmu eV}}\right)^{5/4} 
\\ \!\!  \times   \left(\frac{\mathrm{10^5\,s}}{t}\right)^{1/4} \!\! \times \left(\frac{T_\mathrm{{sys}}}{10\,\mathrm{K}}\right)^{1/2}  \!\! \times \frac{10\,\mathrm{T}}{B_0} \times \left(\frac{\mathrm {0.4\,GeV cm^{-3}}}{\rho_{a}}\right)^{1/2} \, ,
\end{aligned}
\label{Eq.2}
\end{equation}
 with SNR as the signal to noise ratio and $t$ as the total integration time. A forecast of the sensitivity of the DALI apparatuses to Galactic axion DM is shown in Fig. \ref{fig_5}.

\section{Conclusions}
\label{Sec4}
In this work, the quality factor of a dummy dielectric haloscope tuner is measured in the C band, resulting in $Q\simeq 2\times10^3$ for a $\sim$$\lambda/8$ plate distance stack with 20 yttria-stabilized zirconium oxide layers. The power enhancement in $\sim$$\lambda/2$ stacks is expected to be identical, allowing us to probe two resonant frequencies simultaneously in a upcoming implementation \cite{2024JInst..19P1022H}—c.f. Appendix \ref{A1}. From Eq. \ref{Eq.1}, to render the weak axion-induced signal measurable in a reasonable integration time by Fabry-Pérot haloscopes with typical parameters of $A\sim$1 m$^2$ and $B_0\sim$ 10 T, $Q$ factors between $\sim$$10^{3}$ and $\sim$$10^{4}$ are needed at $\sim$100 $\upmu$eV mass to respectively meet the \textit{phenomenologically preferred} axion window \cite{DiLuzio:2016sbl} and the QCD window defined by benchmark axion models and their uncertainties \cite{PhysRevLett.43.103, Shifman1980CanCE, DINE1981199, osti_7063072}. 

It follows from its definition in quantum optics that $Q$ scales linearly with frequency, the permittivity of the dielectric material and the number of layers in series \cite{Millar:2016cjp, millar2018theoretical, De_Miguel_2021, Egge:2020hyo, 2024JInst..19P1022H}. The variation of the electric permittivity and tangential losses of zirconium oxide with frequency from mm to cm wavelengths is negligible, as discussed in Sec. \ref{Sec3}. The argument of $Q\propto \mathrm{N}\varepsilon_{\mathrm{r}}$ for a rising number of layers is also sustained by a small decrease in dielectric losses at lower temperature that allows for a greater number of layers and thus compensates for the decrease of its permittivity by a few percent with temperature, allowing for $Q>2\times10^4$ for $\mathrm{N}\sim50$ in the upper band from 6 to 60 GHz that comprises the range of DALI; while DM detection is enabled by a fainter noise floor at the shorter wavelengths in this band, since the quantum noise is frequency-dependent while electronic devices show a limit of heat dissipation roughly 2–3 times over the quantum noise limit \cite{10.1117/1.JATIS.3.1.014003}. 

The typical scan speed of the DALI haloscopes is of the order of a few GHz/year. As shown in Fig. \ref{fig_5}, this will give access the DALI scientific program to the QCD axion window over the 25--250 $\upmu$eV range in a containable period of time—say, requiring us some 15 years from 25 to $\sim$200 $\upmu$eV; cf. \cite{DeMiguel:2023nmz} for a discussion. Equally encouraging conclusions would be drawn for the search for dark photon DM that is carried out simultaneously with the quest for axions. 

In future work we will introduce the anechoic chamber housing the prototype DALI tuner into the instrument's cryostat to observe the quality factor at the operating temperature of the experiment, since it scales roughly as $\Delta Q / \Delta T \sim 0.1 $ K$^{-1}$ ranging from a few kelvin to room temperature, as deduced from Eq. \ref{Eq3.1}.

\acknowledgments

J.F.H.C. is supported by the Resident Astrophysicist Programme of the Instituto de Astrofísica de Canarias. The work of J.D.M. was supported by RIKEN's program for Special Postdoctoral Researchers (SPDR)—Project Code: 202101061013. We gratefully acknowledge financial support from the Severo Ochoa Program for Technological Projects and Major Surveys 2020-2023 under Grant No. CEX2019-000920-S; Recovery, Transformation and Resiliency Plan of Spanish Government under Grant No. C17.I02.CIENCIA.P5; Operational Program of the European Regional Development Fund (ERDF) under Grant No. EQC2019-006548-P; IAC Plan de Actuación 2022. J.A.R.M. acknowledges financial support from the Spanish Ministry of Science and Innovation (MICINN) under the project PID2020-120514GB-I00. We thank F. Gracia, R. Hoyland, P. Redondo, S. Sordo, and D. Rodríguez for invaluable help. We thankfully acknowledge the technical expertise and assistance provided by the Spanish Supercomputing Network (Red Española de Supercomputación).

\appendix
\section{On resonance over frequency octaves} \label{A1}
If a resonator is composed of two perfect reflectors both with reflectivity $R=1$, light through the layers travels without loss; it performs an infinite number of round trip transits. However, the number of round trip transits is finite if at least one  of the reflectors is a partial reflector with $R<1$. The typical number of round trips ranges from a few to hundreds in Fabry-Pérot resonators \cite{renk2012basics}. Partial reflection at the output coupling mirror corresponds to a reduction of the energy density within the resonator. To know how long does a photon remain in a resonator, we define $\tau_{\mathrm{g}}$ as the average lifetime of the photons, or the decay time of the energy density of radiation in the resonator, that adopt values of the order of a few nanoseconds. Thus, $\tau_{\mathrm{g}}/T$, with $T=2l/c$ being $l\equiv d_{\vartheta}$ the plate distance, is the number of round trips within the resonator. The free spectral range is given by $\Delta \nu = 1/T$. The quality factor of a resonator is equal to 2$\pi$ times the ratio of the energy stored in the resonator and the energy loss per oscillation
period, $Q = \tau_{\mathrm{g}} \omega_0$, where $\omega_0$ is the resonant frequency \cite{renk2012basics}.

As discussed in this work, and also previously in Refs. \cite{DeMiguel:2023nmz, 2024JInst..19P1022H, Hernandez-Cabrera:2024SPIE, Cabrera:2023qkt}, in the DALI interferometers the resonant frequency is tuned by setting a plate distance of a fraction of the scanning wavelength, i.e., $\sim$$\lambda/2$ with a $\sim$$\lambda/(2\sqrt{\varepsilon_{\mathrm{r}}})$ plate thickness for a half-wavelength stack, or a $\sim$$\lambda/8$ spacing with a $\sim$$\lambda/(8\sqrt{\varepsilon_{\mathrm{r}}})$ plate thickness. Interestingly, the one-eighth wavelength stack allows the plate distance to be shortened by about four times without detracting from the signal enhancement. Thus, we use a $\sim$$\lambda/8$ spacing at lower frequencies in order to save room for plate stacking, thereby saving resources while allowing us for the use of a regular solenoid available in the industry rather than a custom magnet with a very large bore; while we employ the $\sim$$\lambda/2$ stack, as employed by other experiments, at higher frequencies. 

\begin{figure}[h]
\centering
\includegraphics[width=.65\textwidth]{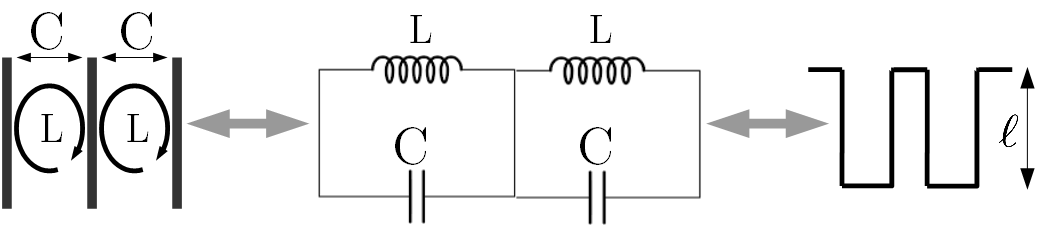}
\caption{A series circuit for a Fabry-Pérot resonator and the performed transformations. Left: the multilayer resonator. Each layer is modeled as a capacitor (C) plus an inductor (L). Middle: LC equivalent circuit with lumped elements. Right: equivalent distributed elements.}
\label{fig_A1}
\end{figure}

Resonance periodicity over $4\lambda$ steps is a fundamental concept that can be also explained with an equivalent electric circuit. This is illustrated in Fig. \ref{fig_A1}. The capacity and inductance of two consecutive plates can be modeled in terms of the plate distance and photon round trip, respectively \cite{Sievenpiper}. Then, the lumped-element equivalent circuit is transformed into a distributed element circuit though Richard's transformations (i.e., opened or shorted transmission lines through Kuroda's identities, with length $\ell$—see \cite{pozar2012microwave, Miguel-Hernández_2019} and references therein. The distributed element circuit is commonly referred to as \textit{commensurate} due to the fact that it is periodic over $4\omega_c$ scales, $4\omega_c$ being the cutoff frequency of the LC filter. This is, Richard's transformations converts frequencies from a $\omega$-domain to a $\Omega$-domain as

\begin{equation}
\Omega=\mathrm{tan}(\beta \ell)=\mathrm{tan}\left(\frac{\omega \ell}{v_p}\right),    
\end{equation}

where $\beta=2\pi/\lambda$ and $\ell$ represents the transmission line length. In the $\Omega$-plane, the functions are periodic with a $\omega \ell/v_p=2\pi$ period. This transformation synthesizes LC-networks using open-circuited (O.C.) and short-circuited (S.C.) transmission line stubs. Once the transformation $\Omega \leftarrow \omega$ is set, we write:
\begin{equation} jX_L=j\Omega L= jL \,\mathrm{tan}(\beta \ell) \, ,
\label{equation_31} \end{equation}  
for the reactance of an inductor, and 
\begin{equation} jB_C=j\Omega C= jC \,\mathrm{tan}(\beta \ell) \, ,
\label{equation_32} \end{equation}  
as the susceptance of a capacitor.

Equation \ref{equation_31} shows that an inductor can be replaced with a short S.C. stub transmission line of length $\beta \ell$ and impedance $L$, while Eq. \ref{equation_32} shows that a lumped capacitor can be replaced with an O.C. transmission line stub of length $\beta \ell$ and characteristic impedance $1/C$.

As mentioned, the domain-transformed low-pass filter has a cutoff frequency at $\omega' =1$. Therefore, at the cutoff frequency we have $\Omega=1=\mathrm{tan}(\beta \ell) \rightarrow \mathrm{tan}(2\pi \ell/\lambda)=1 \rightarrow \ell =\lambda \, \arctan(1) / 2\pi = \lambda/8$. This gives a $\ell= \lambda/8$ equivalent transmission line stub length for $\lambda$ being the wavelength at $\omega_c$. Therefore, at $\omega=2\omega_c$, the lines will have a length of $\ell=\lambda/4$, and because of the same argument the filter response will be periodic in frequency every $4\omega_c$. Finally, we have shown a relationship between the plate distance and the transmission line length $\ell$, where a filtering stage exists below $d_{\vartheta}\sim\lambda/2$, around $\lambda/8$, similarly to  distributed (LC) filters. Above the cutoff frequency, there is a $4\omega_c$ period. On the other hand, the concatenation of LC circuits in series supports the principle that the quality factor, defined as $Q= R \sqrt{C/L}$ with $R$ a resistance, should scale with the number of adjacent layers.



\clearpage
\bibliographystyle{JHEP}
\bibliography{references}




\end{document}